# Exact relativistic 'antigravity' propulsion

F. S. Felber[a)]

*Physics Division, Starmark, Inc., P. O. Box 270710, San Diego, California 92198*

The Schwarzschild solution is used to find the exact relativistic motion of a payload in the gravitational field of a mass moving with constant velocity. At radial approach or recession speeds faster than $3^{-1/2}$ times the speed of light, even a small mass gravitationally repels a payload. At relativistic speeds, a suitable mass can quickly propel a heavy payload from rest nearly to the speed of light with negligible stresses on the payload.

This paper calculates from the exact Schwarzschild solution [1–3] of Einstein's field equation the relativistically exact motion of a payload in the gravitational field of a source moving with constant velocity. In the inertial frame of an observer far from the interaction between the source and payload, the payload motion is calculated exactly for relativistic speeds of both the source and payload and for strong gravitational fields of the source. This paper presents the first *relativistically exact* solution of the unbound orbits of test particles in the *time-dependent* gravitational field of a moving mass.

The relativistically exact bound and unbound orbits of test particles in the strong *static* field of a stationary mass have been thoroughly characterized, for example, in [1–3]. Earlier calculations [1,2,4–6] of the gravitational fields of arbitrarily moving masses were done only to first order in the ratio of source velocity to the speed of light, *c*. Even in a weak static field, these earlier calculations have only solved the geodesic equation for a *nonrelativistic* test particle in the *slow-velocity* limit of source motion. In this slow-velocity limit, the field at a moving test particle has terms that look like the Lorentz field of electromagnetism, called the 'gravimagnetic' or 'gravitomagnetic' field [2,5,6]. Harris [5] derived the *nonrelativistic* equations of motion of a moving test particle in a dynamic field, but only the dynamic field of a *slow-velocity* source.

An exact solution of the field of a relativistic mass is the Kerr solution [1,2,3,7], which is the exact *stationary* (*time-independent*) solution for a rotationally symmetric rotating mass. In the stationary Kerr gravitational field of a spinning mass, the relativistic unbound orbits of test particles have been approximated in [8].

The relativistically exact calculation in this paper shows that a mass radially approaching or receding from a payload with a relative velocity faster than $3^{-1/2}c$ gravitationally repels the payload, as seen by a distant inertial observer. This 'antigravity' is perhaps not so surprising when one considers the following:

(1) The velocity of a particle radially incident on a stationary black hole approaches zero as the particle approaches the event horizon, as seen by a distant inertial observer. Another distant inertial observer would see the same interaction as a black hole approaching the particle, initially at rest, and causing the particle to *accelerate away* from the black hole until the particle attains a speed near the horizon asymptotically approaching the speed of the black hole.

(2) Although time-independent, the Kerr field exhibits an inertial-frame-dragging effect [1,2,6] similar to that contributing to gravitational repulsion at relativistic velocities, namely, a force in the direction of the moving mass. The inertial-dragging force can dominate the radial force of attraction, even near a black hole. Inside the so-called 'static limit' surface of a spinning black hole, an observer can theoretically halt his descent into the black hole, but cannot halt his angular motion induced by inertial-frame dragging [1].

(3) Because the affine connection is *non-positive-definite*, a "general prediction" has been made that general relativity could admit a repulsive force at relativistic speeds [9]. But since the repulsive-force terms are second-order and higher in source velocity, this 'antigravity' at relativistic speeds has not previously been found.

Particularly noteworthy in the new exact solutions is that above a critical velocity any mass, no matter how light or how distant, produces an 'antigravity' field. Though at least twice as strong in the forward direction of motion, the 'antigravity' field even repels particles in the backward direction. This means that a stationary mass will repel masses that are radially receding from it at speeds greater than $3^{-1/2}c$, with obvious cosmological implications.

This paper calculates the exact motion imparted to test particles or payloads by a source moving at constant velocity. A strong gravitational field is not necessary for 'antigravity' propulsion. Solely for the purpose of deriving an exact solution, however, the source is considered to be much more massive than the payload, so that the energy and momentum delivered to the payload have negligible reaction on the source motion.

Figure 1 illustrates the two-step approach of this paper to calculating the exact motion of a payload mass *m* in the field of a source of mass *M* and constant velocity $c\boldsymbol{\beta}_0$. First, the

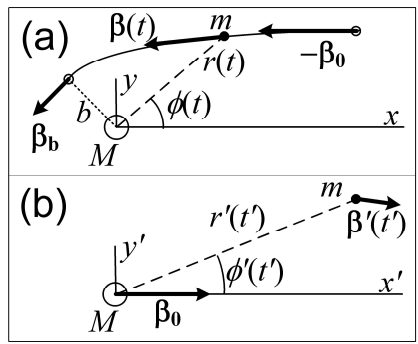

FIG. 1. Two-step exact solution of 'antigravity' propulsion of a payload mass *m* by a relativistic mass *M*: (a) Schwarzschild solution in static field of *M*; and (b) solution Lorentz-transformed to initial rest frame of payload far from *M*. Primes denote Lorentz-transformed quantities.

---
[a)]Electronic mail: starmark@san.rr.com

unique trajectory in the static Schwarzschild field of a stationary mass $M$ is found for which the perigee, the distance of closest approach, of the payload is $b$ and the asymptotic velocity far from the stationary mass is $-c\boldsymbol{\beta_0}$. In the static field of $M$, the trajectory is time-reversible. Second, the trajectory is Lorentz-transformed to a reference frame moving with constant velocity $-c\boldsymbol{\beta_0}$, in which the mass $M$ has a constant velocity $c\boldsymbol{\beta_0}$, and the payload is *initially* at rest. The Lorentz transformation occurs between two inertial observers far from the interaction, in asymptotically flat spacetime.

The exact equations of motion of a ballistic payload in spherical coordinates in the static Schwarzschild field of the mass $M$ are [1–3]

$$dt/d\tau = \gamma_0/\psi \quad , \tag{1}$$

$$(dr/d\tau)^2 + (c^2 + L^2/r^2)\psi = \gamma_0^2 c^2 \quad , \tag{2}$$

$$d\phi/d\tau = L/r^2 \quad , \tag{3}$$

where $\tau$ is the proper time, $\gamma_0 c^2$ is the constant total specific energy, $L$ is the constant specific angular momentum, and $\psi(r) \equiv 1 - 2GM/rc^2$ is the $g_{00}$ component of the Schwarzschild metric.

By substitution of Eq. (1), the equations of motion in coordinate time $t$ become

$$\beta_r^2 = \psi^2 - (1 + L^2/c^2 r^2)\psi^3/\gamma_0^2 \quad , \tag{4}$$

$$\beta_\phi = L\psi/\gamma_0 cr \quad , \tag{5}$$

where $\beta_r \equiv \dot{r}/c$ and $\beta_\phi \equiv r\dot{\phi}/c$ are the $r$ and $\phi$ components of the normalized payload velocity $\boldsymbol{\beta}(r)$ measured by a distant inertial observer. An overdot indicates a derivative with respect to $t$.

As shown in Fig. 1(a), if the payload at perigee, $r = b$, has a speed $b\dot{\phi} = \beta_b c$, then from Eqs. (4) and (5),

$$L = \gamma_0 cb\beta_b/\psi_b \quad , \tag{6}$$

$$\gamma_0^2 = \psi_b^2/(\psi_b - \beta_b^2) \quad , \tag{7}$$

where $\psi_b \equiv \psi(b)$. And as shown in Fig. 1(a), if the payload has a speed $\beta_0 c$ far from the mass $M$, where $\psi \approx 1$ and $\dot{r} \gg r\dot{\phi}$, then from Eq. (4),

$$\gamma_0^2 = 1/(1 - \beta_0^2) \quad . \tag{8}$$

The payload speeds far from the mass $M$ and at perigee are related through Eqs. (7) and (8) by

$$\beta_b^2 = \psi_b - \psi_b^2(1 - \beta_0^2) \quad . \tag{9}$$

Since the payload is moving in the static gravitational potential of the mass $M$ in this reference frame, the payload speed given by Eqs. (4) and (5) is a function of $r$ only,

$$\beta(r) = \psi[1 - (\psi/\gamma_0^2) + (L/\gamma_0 cr)^2(1-\psi)]^{1/2} \quad . \tag{10}$$

And $\phi(r)$ is found by integrating the exact orbital equation in a Schwarzschild field [3],

$$(d\rho/d\phi)^2 + \rho^2 = 2GM(\rho^3/c^2 + \rho/L^2) + (c\beta_0\gamma_0/L)^2 \quad , \tag{11}$$

where $\rho \equiv 1/r$.

The radial acceleration of the payload,

$$c\dot{\beta}_r = \frac{-GM}{r^2}\left(\frac{3\psi^2}{\gamma_0^2} - 2\psi\right) + \frac{L^2\psi^2}{\gamma_0^2 r^3}\left(\psi - \frac{3GM}{rc^2}\right) \quad , \tag{12}$$

indicates repulsion by the stationary mass $M$ whenever

$$\gamma_0^2 > \frac{3}{2}\psi\left[1 - \frac{L^2}{GMr}\left(\frac{\psi}{3} - \frac{GM}{rc^2}\right)\right] \quad . \tag{13}$$

Equation (13) is the exact relativistic strong-field condition for 'antigravity' repulsion of a payload to be measured by a distant inertial observer. A payload far from the stationary mass $M$, for both $r \gg b$ and $r \gg GM/c^2$, is seen to be repelled by $M$ whenever $\gamma_0^2 > 3/2$ or $\beta_0 > 3^{-1/2}$.

Next, we transform to the inertial reference frame shown in Fig. 1(b), in which the mass $M$ moves in the $x$ direction at constant speed $\beta_0 c$, and the mass $m$ is initially at rest at $r'_0 \equiv r'(0)$ and $\phi'_0 \equiv \phi'(0)$. (Since only unbound orbits are considered here, $r'_0$ can always be chosen large enough that $\phi'_0$ is negligible.) In the rest frame of $M$, the $x$ and $y$ components of $\boldsymbol{\beta}$ were

$$\beta_x = -\beta_r \cos\phi - \beta_\phi \sin\phi \quad , \tag{14}$$

$$\beta_y = -\beta_r \sin\phi + \beta_\phi \cos\phi \quad . \tag{15}$$

In the Lorentz-transformed frame, in which $M$ moves in the $x$ direction at constant speed $\beta_0 c$, the components of the exact payload velocity $c\boldsymbol{\beta}'$ and acceleration $cd\boldsymbol{\beta}'/dt'$ are given by [10]

$$\beta'_x = (\beta_0 + \beta_x)/(1 + \beta_0\beta_x) \quad , \tag{16}$$

$$\beta'_y = (\beta_y/\gamma_0)/(1 + \beta_0\beta_x) \quad , \tag{17}$$

$$\frac{d\beta'_x}{dt'} = \frac{\dot{\beta}_x}{\gamma_0^3(1 + \beta_0\beta_x)} \quad , \tag{18}$$

$$\frac{d\beta'_y}{dt'} = \frac{(1 + \beta_0\beta_x)\dot{\beta}_y - \beta_0\beta_y\dot{\beta}_x}{\gamma_0^2(1 + \beta_0\beta_x)^3} \quad . \tag{19}$$

Equations (4), (5), and (14) – (17) completely define the exact payload velocity measured by a distant inertial observer, to whom the mass $M$ appears to be moving in the $x$ direction at constant speed $\beta_0 c$. Two cases of special interest for their simplicity are briefly analyzed here: (1) The exact payload velocity with purely radial motion ($L = 0$) in the strong field of a black hole; and (2) the approximate velocity with arbitrary unbounded motion in a weak field.

For purely radial motion, Eqs. (4) and (14) give the exact speed of a payload in the field of a moving black hole as

$$\beta' = \frac{\beta_0 - (\Psi^2 - \Psi^3/\gamma_0^2)^{1/2}}{1 - \beta_0(\Psi^2 - \Psi^3/\gamma_0^2)^{1/2}} \quad , \tag{20}$$

where $\Psi \equiv 1 - r_{BH}/r$, and $r_{BH}$ is the radius of the black hole. Figure 2 shows this radial velocity of the payload for several values of the black hole speed, $\beta_0$.

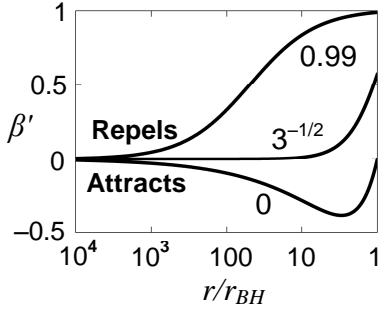

FIG. 2. Exact payload speed vs. normalized distance to black hole for constant black-hole approach speeds $\beta_0$ indicated. All payloads start from rest at $r = 10^4 r_{BH}$.

Figure 2 illustrates several interesting results concerning the observations of a distant inertial observer witnessing a black hole directly approaching a payload. At any closing speed, the final payload speed is always the speed of the black hole, approached as the payload approaches $r_{BH}$. At a closing speed faster than $3^{-1/2}c$, the payload will be continuously repelled by the black hole at any distance. At slower closing speeds, the payload will always be seen to be repelled by the strong field of the black hole at least within a radius $3r_{BH}$. Equation (18) shows that Eq. (13) is the exact relativistic strong-field condition for 'antigravity' repulsion of a payload measured by a distant observer in either inertial reference frame. That is, either observer will see the payload repelled when $\beta_0 > (1 - 2/3\psi)^{1/2}$.

For any unbounded motion of a payload about a stationary mass, a normalized gravitational potential is defined as $V(r) \equiv (\gamma_0 - \gamma)/(1 - \psi_b)$, where $\gamma(r) \equiv (1-\beta^2)^{-1/2}$. This exact potential is the payload total specific energy, less the specific rest plus kinetic energies, normalized to –2 times the classical potential at perigee. In the weak-field approximation, this normalized potential, $V \approx -(\gamma_0^3/2R)(1-3\beta_0^2 + \beta_0^2/R^2)$, plotted against $R \equiv r/b$ in Fig. 3, is independent of field strength and of angular momentum. The dashed curve is the normalized critical radius, $R_c = (1-1/3\beta_0^2)^{-1/2}$, beyond which the potential is repulsive ($dV/dr < 0$).

By combining Eqs. (4), (12), and (18) with the FitzGerald-Lorentz contraction [10], $x = s'/\gamma_0$, where $s' \equiv x' - \beta_0 ct'$, the weak field of mass $M$ directly approaching a payload, as measured by a distant inertial observer in the *initial* rest frame of the payload (when it is far from $M$), is found to be

$$cd\beta_x'/dt' \approx -\gamma_0(1-3\beta_0^2)GM/s'^2 \quad . \tag{21}$$

This same result can be derived from the geodesic equation. On the $x$ axis, the weak-field metric of mass $M$ moving with constant speed $c\beta_0$ along the $x$ axis is linearized as $g_{\mu\nu} = \text{diag}(1,-1,-1,-1) + h_{\mu\nu}$. The nonzero components of $h_{\mu\nu}$ in Cartesian coordinates, $(ct', x', y', z')$, are $h_{00} = h_{11} = 2(1+\beta_0^2)\Phi$, $h_{01} = h_{10} = -4\beta_0\Phi$, and $h_{22} = h_{33} = 2\Phi/\gamma_0^2$, where the dimensionless potential, $\Phi \equiv -GM\gamma_0/c^2s'$, satisfies the harmonic gauge condition, $\partial\Phi/\partial t' + c\beta_0\partial\Phi/\partial x' = 0$. From the geodesic equation, the equation of motion for the payload in the weak field of $M$ is $d^2s'/d\tau^2 + (c^2/\gamma_0^4)d\Phi/ds' \approx 0$. The first integrals of the motion are

$$dx'/d\tau \approx c\beta_0(1-3\beta_0^2)\Phi \quad , \tag{22}$$

$$dt'/d\tau \approx 1 - (1+\beta_0^2)\Phi \quad . \tag{23}$$

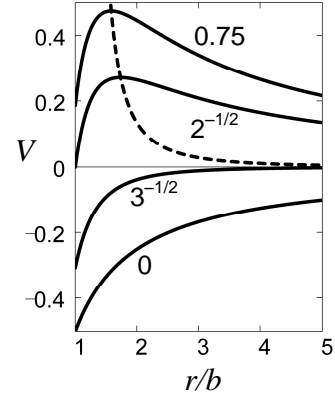

FIG. 3. Normalized potential of weak gravitational field of stationary mass vs. normalized distance for values of $\beta_0$ indicated. Above dashed curve ($r/b = R_c$), potential is repulsive ($dV/dr < 0$).

In the weak-field approximation, terms of order $\Phi^2$ are neglected, so that $dx'/d\tau \approx dx'/dt'$, and the acceleration of the payload from Eq. (22) is $d^2x'/dt'^2 \approx -\gamma_0(1-3\beta_0^2)GM/s'^2$, in agreement with Eq. (21).

The separation $\{R'\}_{\text{ret}}$ of the payload and a directly approaching mass $M$ at the retarded time $t' - \{R'\}_{\text{ret}}/c$ is $\{R'\}_{\text{ret}} = s'/(1-\beta_0)$. In terms of this retarded separation, the acceleration of the payload from Eq. (21) is

$$cd\beta_x'/dt' \approx -\gamma_0^5(1+\beta_0)^2(1-3\beta_0^2)GM/\{R'^2\}_{\text{ret}} \quad , \tag{24}$$

in agreement with the weak retarded field found by Liénard-Wiechert methods [11].

In the weak-field approximation, the field on a stationary test particle is the same as the field on a payload moving freely along a geodesic, as long as the payload starts from rest. That is, the 'gravitomagnetic' terms in the geodesic equation are of the same order as terms that are neglected in the weak-field approximation.

From Eq. (10), the minimum speed reached by the payload in Fig. 1(a), corresponding to the maximum deceleration of the payload by the weak static field of $M$, is

$$\beta_{\min} \approx \beta_0[1 - (2GM/bc^2)/R_c^2] \quad , \tag{25}$$

at the normalized critical radius, $R_c$. From Eqs. (16) and (25), in the Lorentz-transformed frame of Fig. 1(b), the maximum speed delivered to a payload, initially at rest, is

$$\beta_{\max}' \approx (2GM/bc^2)\gamma_0^2(\beta_0 - 1/3\beta_0) \quad . \tag{26}$$

This field of a radially approaching mass is greater than that of a radially receding mass by a factor of about $\gamma_0^2(1+\beta_0^2)$, according to Eq. (18).

The weak-field condition used to derive Eq. (26) is $(2GM/bc^2)\gamma_0^2 \ll 1$. Therefore, we find that the maximum speed that can be delivered to a payload, initially at rest, by the *weak field* of a much heavier mass moving at constant speed $\beta_0 > 3^{-1/2}$ is $\beta_{\max}' \ll \beta_0 - 1/3\beta_0$. As was seen in Fig. 2, the maximum speed that can be delivered to a payload, initially at rest, by the *strong field* of a black hole directly incident on it at any speed $\beta_0$ is $\beta'_{max} = \beta_0$.

Whether the payload is accelerated by a strong or a weak field, the payload travels along a geodesic. The only stresses on the payload, therefore, are the result of tidal forces in the accelerated frame of the payload. These stresses can be arranged by choice of the trajectory to be kept within acceptable limits. Greater practical problems for gravitational propulsion are finding a suitable and accessible driver mass at relativistic velocities, and maneuvering the payload in and out of the driver trajectory.

The seeming scarcity of suitable relativistic drivers in our galactic neighborhood may well be due to drag by just the sort of gravitational repulsion analyzed in this paper. The analysis found that at radial approach or recession speeds faster than $3^{-1/2}$ times the speed of light, any mass gravitationally repels a payload at any distance. The forward 'antigravity' field of a suitably heavy and fast mass might be used to propel a payload from rest to relativistic speeds.